\documentclass[12pt]{amsart}

\pdfoutput=1

\usepackage[text={500pt,650pt},centering]{geometry}
\usepackage{bbm}
\usepackage{enumitem}
\usepackage{url}
\usepackage{amssymb,amsfonts,amsmath,amsthm}
\usepackage{esint}
\usepackage{mathtools} 
\usepackage{graphicx}
\usepackage{MnSymbol}
\usepackage{mathtools} 
\usepackage[bottom]{footmisc}
\usepackage[colorlinks=true, pdfstartview=FitV, linkcolor=blue, citecolor=magenta, urlcolor=blue,pagebackref=false]{hyperref}
\usepackage{microtype}
\usepackage{cite}
\usepackage{xcolor}
\usepackage{bigints}

\usepackage{bm}
\usepackage{dsfont} 
\usepackage{notations} 
\usepackage{mathrsfs} 

\newenvironment{talign*}
 {\csname align*\endcsname}
 {\endalign}

\begin{document}

\title[Statistical inference and high-dimensional statistics]{The mighty force: Statistical inference \\and high-dimensional statistics}

\author[E. Aurell]{Erik Aurell}
\address[Erik Aurell]{Department of Computational Science and Technology, AlbaNova University Center, SE-106 91 Stockholm, Sweden.}
\email{eaurell@kth.se}

\author[J. Barbier]{Jean Barbier}
\address[Jean Barbier]{International Center for Theoretical Physics, Trieste, Italy.}
\email{jbarbier@ictp.it}

\author[A. Decelle]{Aurelien Decelle}
\address[Aurelien Decelle]{Departamento de Física T\'eorica, Universidad Complutense de Madrid, 28040 Madrid, Spain.}
\email{adecelle@ucm.es}

\author[R. Mulet]{Roberto Mulet}
\address[Roberto Mulet]{Group of Complex Systems and Statistical Physics, Department of Theoretical Physics, Physics Faculty, University of Havana, Cuba.}
\email{roberto.mulet@gmail.com}

\date{}
\maketitle

\begin{quotation}
\begin{flushright}
Full well hath Clifford play'd the orator,\\
Inferring arguments of mighty force.\\
%But, Clifford, tell me, didst thou never hear\\
%That things ill-got had ever bad success?\\
%And happy always was it for that son\\
%Whose father for his hoarding went to hell?\\
%I'll leave my son my virtuous deeds behind;\\
%And would my father had left me no more!\\
%For all the rest is held at such a rate\\
%As brings a thousand-fold more care to keep\\
%Than in possession and jot of pleasure.\\
%Ah, cousin York! would thy best friends did know\\
%How it doth grieve me that thy head is here!\\
--- \textit{Henry VI}, Part 3, Act II, Scene II.
\end{flushright}
\end{quotation}

\section{Introduction}\label{sec21.1}

\textit{Inference} is an English noun formed on the verb \textit{infer}, from the Latin
\textit{inferre}, meaning to carry (fero) in or into (in-) something.
That originally concrete %and active 
meaning can still be felt in 
the portal quote of this chapter\footnote{To appear as a contribution to the edited volume "Spin Glass Theory \& Far Beyond - Replica Symmetry Breaking after 40 Years", World Scientific.}.
%a quote adduced in Webster's Dictionary \cite{1913Webster}
%     \begin{quotation}
%     \begin{center}
%     Full well hath Clifford played the orator,\\
%     Inferring arguments of mighty force.  \end{center}  
%     \end{quotation}
In modern non-technical use the meaning of inference 
is more abstract, and rendered either as 
``A conclusion reached on the basis of evidence and reasoning''
or as ``The process of reaching such a conclusion'' \cite{wikipedia-inference}.
In scientific language these translate into characteristics of a phenomenon
that are not observed directly, but which are arrived at (inferred) from observations
with the help of mathematical and/or statistical methods, and those methods themselves.
We will discuss three prominent examples of inference in both
senses of modern usage, and how they
naturally open up new perspectives and possibilities.
%the Far Beyond of this chapter.

Statistical physics is played out on the terrain between individual items and distributions over
properties of items. The canonical example is the Langevin equation which describes the motion
of a Brownian particle interacting with a thermal reservoir, and the Fokker-Planck equation which describes the evolution of the distribution of possible positions and velocities of the particle.
In inference the goal can analogously be to reach one conclusion or retrieve one object, or to establish characteristics of a distribution over objects. 
The second kind of inference is also called \textit{statistical inference}.
We will here discuss inference in this sense.

In the ``big-data era'', statistical inference of different kinds is routinely performed based on data sets containing millions or even billions of samples, which themselves may live in spaces of tremendously large dimensionality. In this realm, classical statistical wisdom and tools fail: new mathematics and algorithms able to tackle the phenomena emerging in this regime are necessary. In the very same way, phase transitions were understood to emerge from the complexity (\textit{i.e.} high-dimensionality) of physical systems more than a century ago, whose understanding required to develop statistical mechanics. It turns out that this is more than an analogy as the theory and methods to perform \emph{high-dimensional inference} are directly connected to statistical mechanics as we will see in this chapter (and others in the book \cite{RSB}). High-dimensional inference itself is part of a broader statistical theory of complex systems, referred to as \emph{high-dimensional statistics}, a very active research field at the crossroads of (statistical) physics, computer science, information theory and machine learning, and which is the powerhouse of modern information processing systems.

%A convenient way to represent the main objects of study in high-dimensional statistical problems, namely probability distributions describing (very) many interacting variables of all sorts, is using graphs and hypergraphs. A graph consists of nodes and links between them representing pair-wise dependencies. More generically, in a \emph{hypergraph}, hyperlinks between more than two nodes represent higher than pair-wise dependencies. It is often the case that (hyper) graphs are identified with the probability distribution they represent. Questions of interests related to these distributions, such as performing statistical inferences, are then naturally expressed using graph-based quantities as it will become clear later on.
%It is well known that hyper-graphs are equivalent to graphs by introducing dummy variables and constraints, and also that hypergraphs often are a more convenient representation of a given problem. 
%A prominent class of hypergraphs are those representing instances of constraint satisfaction problems (or ``combinatorial optimization''), where hyperlinks are the hard constraints that must verify the solutions of the problem, and the probability distribution in this case is just uniformly weighing the solutions (if any) verifying all the constraints simultaneously. For simplicity we will in this chapter refer to all such representations as graphs.
%A probability distribution is then more fully represented by a graph, and numbers associated to the nodes and links of the graph. 

Let us now dive in some specific models whose richness, generality and wide applicability make them ideal candidates to showcase modern high-dimensional inference and its links with physics. A canonical example of statistical inference is to retrieve the interaction graph and the parameters of the interactions of an energy function of the Ising or Potts model type. In statistics the first task would be called \textit{model learning} and the second \textit{parameter inference}.
All together the task would be referred to as \textit{learning and inference in an exponential family}~\cite{WainwrightJordan2008}.
The set of Gibbs-Boltzmann distributions with unknown energy function is the model family, which in statistics is called exponential because the energy function appears in the exponent. In statistical physics the term \textit{inverse Ising/Potts model} has been used~\cite{ChauNguyen2017}. %posited on that the direct Ising / Potts problem is to compute means and correlations from the model parameters. 
We prefer to avoid this term since it is now well established, and discussed in detail in~\cite{ChauNguyen2017}, that the best learning and inference procedure is not (or is rarely) to infer model parameters from means and correlations. We will instead use the term \textit{Direct Coupling Analysis} (DCA) introduced in~\cite{Weigt2009}
which does not have the same restrictive connotation, and which 
encompasses several new aspects which have turned out to be very important in applications.
Below we will discuss the distinguishing characteristics of DCA, its main methods applied, and successes in biological data analysis.

After DCA %namely inference \emph{of} a graph and its parameters, 
we will move towards the problem of \emph{community detection} (COMDEC) --- or graph partitioning ---, the paradigmatic model of inference from an observed graph. 
%COMDEC is a more specific problem than DCA, but has lead to the understanding of many crucial phenomena such as the information-theoretic threshold of the detectable/undetectable phases and its relation to other combinatorial optimization problems such as coloring. 
The main focus of COMDEC is to partition a given graph into two or more communities, possibly without prior knowledge on the number of hidden communities or their statistical properties, such as the connectivity among the different communities or their fractions of nodes. In this model, whose study has quickly become a whole field of research \cite{abbe2017community,moore2017computer}, one notable success of statistical physics has been to exhibit in the celebrated \emph{Stochastic Block Model (SBM)} the notion of phase where a given graph could not be distinguished by any mean from a purely random structure-less graph despite the fact that it has been built according to a process dependent on hidden communities (\textit{i.e.} sets of nodes sharing the same property). Of course, this is true in the thermodynamic limit where the number of nodes diverges. Digging deeper in the landscape of solutions of the problem, it has been then demonstrated the presence of interesting regimes such as when the true partition can be recovered information-theoretically, but is impossible in practice to retrieve without prior knowledge \cite{decelle2011asymptotic}. Other approaches from the physics community \cite{zhang2014scalable} have also emphasized the presence of retrieval and spin glass phases, suggesting that the design of new methods for clustering should be done with great care. We refer to \cite{RSB} for a discussion of phase transitions in high-dimensional inference and learning and in combinatorial optimization. 

The Gibbs-Boltzmann distribution which underlies both DCA and COMDEC
is the thermodynamic equilibrium
of a system which exchanges energy with a thermal reservoir. In the weak-interaction limit
this distribution is the fixed point of a stochastic process which satisfies detailed
balance, \textit{e.g.} the Metropolis-Hastings process. 
The parameters of the distribution are then related to the 
process rates and can be assimilated to causes,
in opposition to correlations, which can be assimilated to
effects\footnote{This relation is the likely backdrop to ``Direct Coupling'' in the acronym DCA.}. 
Non-equilibrium dynamics which do not obey detailed balance on the other hand
typically lead to distributions qualitatively different
from Gibbs-Boltzmann, for a famous instance, see \cite{Derrida2007}. 
As a third contribution of statistical physics to statistical inference
we will review attempts to extend the cavity method in order 
to describe data which is not generated
by a process in detailed balance, and/or where one has access to a time series.
This \emph{dynamic cavity method} (DYNCAV) brings specific challenges compared to standard (static) cavity, which 
we will outline together with some advances and current problems.

To sum up, the purpose of the present chapter is to showcase a selection of contributions from the spin glass community at large to high-dimensional statistics, by focusing on three important  ``graph-based'' models and methodologies having deeply impacted the field: inference \emph{of} graphs (DCA), inference \emph{from} graphs (COMDEC), and inference \emph{from graphs encoding causal relations}, which is one of the motivations of (DYNCAV).

\section{Direct Coupling Analysis (DCA)}\label{sec21.2}

\subsection{Definition, under-sampling, and evaluation criteria }
\label{sec21.2.1}
In this chapter we use DCA
as the collective term encompassing 
a wide array of methods
to arrive at point estimates of
parameters in 
the exponential family of the Gibbs-Boltzmann distribution
of Ising or Potts models.
Point estimates means that the outcome of the analysis is
one numerical value for each parameter. Ising and Potts 
models have pairwise Hamiltonians, which 
for Ising models are over spins ($s_i=\pm 1$)
\begin{equation}\label{eIH}
    \mathscr{H}_{\mathbf{J},\mathbf{h}}(\mathbf{s}) = -\sum_i h_is_i - \sum_{i<j}J_{ij}s_is_j.
\end{equation}
Up to a re-definition of the 
parameters, spin variables are equivalent to Boolean
variables $b_i\in \{0,1\}$. Potts models are analogously built on 
categorical variables $x_i\in \{1,2,\ldots,q\}$.

In equilibrium statistical mechanics the probability of a configuration $\mathbf{s}$ is the Gibbs-Boltzmann distribution
\begin{equation}\label{eIBD}
    p_{\mathbf{J},\mathbf{h}}(\mathbf{s})=\frac{1}{\mathcal{Z}(T,\mathbf{J},\mathbf{h})}\exp\Big(-\frac{1}{k_B T} \mathscr{H}_{\mathbf{J},\mathbf{h}}(\mathbf{s})\Big).
\end{equation}
In inference temperature is not relevant as it can always be absorbed in a redefinition of the parameters of the Hamiltonian. We will therefore from now on set $k_B T=1$. The normalization $\mathcal{Z}$, the partition function, is then a functional of the Hamiltonian only, and given by
\begin{equation}\label{ePFIIP}
    \mathcal{Z}(\mathbf{J},\mathbf{h}) = \sum_{\mathbf{s}} \exp\Big(-\mathscr{H}_{\mathbf{J},\mathbf{h}}(\mathbf{s}) \Big).
\end{equation}
%We now state the following definition:

\vspace{5pt}
\noindent \textbf{DCA, the basic task}: \emph{Let $D=\{\mathbf{s}^{(m)}\}_{m\le M}$ a set of independent samples of configurations of an Ising or Potts system, the parameters of which are unknown. The basic task of DCA is to infer the parameters from $D$.}
\vspace{5pt}
%The inference problem could be formulated starting from the traditional quantities of statistical mechanics, above all single-spin magnetizations and spin-spin pair-wise correlations. In this sense DCA is the opposite of the Ising problem or the Potts problem, and can therefore be called \textit{inverse Ising problem} or \textit{inverse Potts problem}. There is a sense in which this terminology is correct, since maximum likelihood inference only needs those traditional quantities. We discuss this issue in Sect.~\ref{sec21.2.2}.

Obviously one could relax the assumption that the samples are independent. In most applications where DCA has been used successfully, samples have almost surely not been independent. However, dependence between samples is a complication and has been little investigated in the methodological literature, except under the assumption that they have been generated by a dynamics of known type (Glauber model~\cite{Bresler2018,Dutt2021} or parallel dynamics~\cite{roudi2011mean}). On the other hand, in this setting other methods are available which take advantage of access to time series data \textit{cf.}~\cite{Zeng2013}.
We will therefore throughout assume that samples are independent draws from the same probability distribution. 

The number of parameters of a general Ising model on $N$ spins is $N(N+1)/2$, and correspondingly larger for a Potts model. Each of the $M$ samples consists of $N$ Boolean variables (categorical variables for the Potts model).
It is easy to imagine that for many data sources, $M$ can be of the order of $N$, if not less. 
Such problems are \textit{under-sampled}.
Indeed, in the flagship application of DCA to biological data 
reviewed below in Sect.~\ref{sec21.2.5}, a protein typically consists of $10^2-10^3$ amino acids ($N$), a protein family would often consist of
$10^4-10^5$ proteins ($M$), and  
there are $20$ amino acids ($q$),
hence $\alpha\equiv q M N/(q^2 N^2)\approx 1$.
We find this sufficiently important to state it as 

\vspace{5pt}
\noindent  \textbf{DCA, distinguishing property}: \emph{
Real-world applications of DCA are usually under-sampled: the amount of data
is of the order of the number of parameters, and sometimes less. Any version of DCA intended for practical use hence has to be evaluated in the under-sampled regime.}
\vspace{5pt}

\noindent If the number of samples $M$ is less than the number of parameters $N$, it is not possible to infer all parameters. A natural assumption is to then use supplementary conditions. \textit{Sparsity} in general refers to the situation where a fraction of all parameters are known to be zero; if 
the remaining fraction of non-zero parameters is small enough, it is (in principle)
possible to infer them.
%The problem when it is known that a fraction of parameters are zero, but it is not known which parameters are zero, has been widely studied in statistics, particularly for the linear problem known as \textit{compressed sensing}. One standard (though not optimal) approach is to introduce an $L_1$ penality function on parameters. Alternatively, an inference problem with sparsity may be considered as mostly a model learning problem, where the difficult part is to determine the graph representing non-zero parameters. 
We here only state the empirical fact that sparsity assumptions and $L_1$ penalties have (up to now) not been very useful in DCA applications.
Usually a much simpler $L_2$ penalty (which does not enforce sparsity)
has given more useful predictions.
We posit that the reason is the following:

\vspace{5pt}
\noindent \textbf{DCA, importance of criteria}: \emph{
In real-world applications of DCA one is rarely or never interested in
all the parameters. In practice, one is interested in a small subset of leading 
predictions, typically the inferred parameters of largest numerical value.
}

\vspace{5pt}
\noindent We note that if the task is from the outset formulated as retrieving
$k$ largest parameters where $k$ is much less than $N^2$ and also less than $M$,
the problem is no longer under-sampled. All successful applications of DCA 
known to us are in fact of this type. 
%We will below review more precise statements that have been arrived at for specific variants of DCA that are also computationally efficient, and can be used on large problem instances.
The task of inferring  
$k$ largest parameters in the Hamiltonian from $M$ samples with some given inference
procedure is a mathematically well-defined problem. 
Some (not all) of the procedures to be reviewed below are
\textit{statistically consistent}, meaning that given 
an infinite number of samples they will almost surely return the correct value.
A problem which has been much less studied is the spread around
the correct value given finite number of samples.
We formulate this as

\vspace{5pt}
\noindent \textbf{DCA, a problem in extreme value statistics}: \emph{
Given $M$ independent samples from a Gibbs-Boltzmann distribution of
an Ising/Potts model and a DCA procedure, find the probability distribution of the $k$
largest inferred model parameters.
}

\vspace{5pt}
An obvious analogy showing that the problem is not trivial
is the Marchenko-Pastur distribution of empirical (sample) correlations
from a distribution with given ensemble (model) correlations\footnote{Marchenko-Pastur relates the sample correlation matrix $\mathbf{C}^*$ to the ensemble correlation $\mathbf{C}$ of the underlying probability law. 
For models of the Ising/Potts type $\mathbf{C}$ is a function of the parameters $\mathbf{J}$
(forward Ising/Potts problem). At the same time, some versions of DCA 
infer parameters $\mathbf{J}^*$ from $\mathbf{C}^*$ (inverse Ising/Potts problem solved by naive mean-field or TAP).
For these versions of DCA Marchenko-Pastur hence implies a complicated  
but in principle precise probabilistic relation between $\mathbf{J}$ and
$\mathbf{J}^*$.  
More generally, for versions of DCA that do not rely on correlations, e.g. pseudo-likelihood maximization,
there should be an analogous though so far imperfectly known probabilistic relation between 
$\mathbf{J}$ and
$\mathbf{J}^*$ which does not go through ensemble and sample correlations.
}
\cite{marchenko1967distribution}. An equally obvious
complication is that computing correlations is a linear transformation 
of the data\footnote{For simplicity, assume zero means.}, while all
DCA procedures are nonlinear transformations.
For the pseudo-likelihood maximization method (PLM, see below)
and assuming that the Ising coupling parameters $J_{ij}$
are independent random variables of typical amplitude $N^{-\frac{1}{2}}$,
the expected mean square error of the inferred parameters 
was computed in \cite{Berg-2017a} and \cite{Bachschmid-Romano2017} using 
the replica and cavity methods.
More recently this analysis was extended to the dilute case
\cite{Abbara2019}, 
\textit{i.e.} when the known interaction matrix is locally tree-like,
again with the use of replica and cavity methods.

The full probability distribution of the retrieved parameters has 
to our knowledge only been considered theoretically in \cite{Xu2018},
for $J_{ij}$ distributed as in the Sherrington-Kirkpatrick model,
and for an $L_2$-regularized 
naive mean-field inference,  a DCA procedure which can be implemented as a nonlinear
matrix transformation \cite{Andreatta-2014a}.
Some numerical results for other and more uneven distributions of
interaction parameters were reported in \cite{Gao2018}.
We believe that the issue merits further attention.

\subsection{Thermodynamics, maximum likelihood, and max-entropy}
\label{sec21.2.2}
As we have defined it above, DCA has aspects not germane to
equilibrium statistical mechanics. Nevertheless, a relation obviously
exists, and has been the source of inspiration for several important
versions of DCA.
Let us start from the Helmholtz free energy at a conventional value
of $k_B T=1$:
\begin{equation}\label{eHFE}
    \mathcal{F}(\mathbf{J},\mathbf{h}) = -\log \mathcal{Z}(\mathbf{J},\mathbf{h}) ,
\end{equation}
In the forward problem, the first and second order moments 
are given by derivatives of (\ref{eHFE}):
\begin{equation}
    \chi_i = \left<s_i\right> = - \frac{\partial \mathcal{F}}{\partial h_i}(\mathbf{J},\mathbf{h}), \qquad  \phi_{ij} = \left<s_is_j\right> =
    - \frac{\partial \mathcal{F}}{\partial J_{ij}}(\mathbf{J},\mathbf{h}) .
\end{equation}
The correlation (covariance matrix) is defined by $\chi_{ij}=\phi_{ij}-\chi_i\chi_j$ 
and can also be expressed as
\begin{equation}
%\chi_{ij} = -\frac{\partial \mathcal{F}}{\partial h_i}(\mathbf{J},\mathbf{h})\frac{\partial \mathcal{F}}{\partial h_j}(\mathbf{J},\mathbf{h}) + \frac{\partial^2 \mathcal{F}}{\partial h_i\partial h_j}(\mathbf{J},\mathbf{h}) . 
 \chi_{ij} = - \frac{\partial^2 \mathcal{F}}{\partial h_i\partial h_j}(\mathbf{J},\mathbf{h}) .
\end{equation}
In inference the roles of the parameters $\mathbf{J}, \mathbf{h}$ and the observables $\boldsymbol{\phi}, \boldsymbol{\chi}$ are reversed: the latter are now fixed, the former to be determined. 
As noted in \cite{Sessak-2009a} one can construct a thermodynamic potential
which is a function of the observables, the partial derivatives of which give the parameters. That potential is a Legendre transform of the Helmholtz free energy with respect to both couplings and fields:
\begin{equation}\label{eELT}
    \mathcal{S}(\boldsymbol{\phi},\boldsymbol{\chi}) = \min_{\mathbf{J},\mathbf{h}}\Big[ -\sum_ih_i\chi_i - \sum_{i<j}J_{ij}\phi_{ij} - \mathcal{F}(\mathbf{J},\mathbf{h}) \Big].
\end{equation}
It is immediate that this is the entropy of the  Gibbs-Boltzmann distribution \eqref{eIBD} as function of 
magnetizations and correlations.
Given that $\mathcal{S}$ is a convex function, the Helmholtz free energy is given by a second Legendre function 
of the entropy functional \textit{i.e.}

\begin{equation}
    \mathcal{F}(\mathbf{J},\mathbf{h}) = \min_{\boldsymbol{\phi},\boldsymbol{\chi}}\Big[ -\sum_ih_i\chi_i - \sum_{i<j}J_{ij}\phi_{ij} - S(\boldsymbol{\phi},\boldsymbol{\chi}) \Big].
\end{equation}
From this follows a second type of variational equations:
\begin{equation}
    J_{ij} = -\frac{\partial \mathcal{S}}{\partial \phi_{ij}}(\boldsymbol{\phi},\boldsymbol{\chi}),\qquad h_{i} = -\frac{\partial \mathcal{S}}{\partial \chi_{i}}(\boldsymbol{\phi},\boldsymbol{\chi}).
\end{equation}
Let us now assume $M$ independent samples and 
the maximum likelihood criterion to infer parameters
from samples. 
According to this criterion the best estimate $\theta^{ML}$ based on the available data is given by 
\begin{equation}\label{eMLES}
    \theta^{ML}=\hbox{arg max}_{\theta} \,\, p(x_1,\dots, x_M \mid \theta).
\end{equation}
For numerical reasons,
it is common practice to maximize the average logarithm of the likelihood
\begin{equation}
\mathscr{L}_D(\mathbf{J},\mathbf{h})=\frac{1}{M}\log p(x_1,\dots, x_M \mid \mathbf{J},\mathbf{h}) .
\end{equation}
A penalty (for concreteness here $L_2$ penalty) in
maximizing log-likelihood means to maximize instead
\begin{equation}
\mathscr{L}_D(\mathbf{J},\mathbf{h})_{\boldsymbol{\lambda}}=\frac{1}{M}\log p(D \mid \mathbf{J},\mathbf{h}) 
- \sum_{i} \lambda_{i} |h_i|^2 - \sum_{ij} \lambda_{ij} |J_{ij}|^2 .
\end{equation}
Clearly this is the same as maximum likelihood with a
modified probability distribution
\begin{equation}\label{eq:regularized-Gibbs-Boltzmann}
    p_{\boldsymbol{\lambda},D}(\mathbf{J},\mathbf{h})
    \sim \prod_{i=1}^M \exp\Big({-\mathscr{H}_{\mathbf{J},\mathbf{h}}(\mathbf{s}^{(i)})- \sum_{i} \lambda_{i} |h_i|^2 - \sum_{ij} \lambda_{ij} |J_{ij}|^2}\Big).
\end{equation}
For Ising parameters 
$\mathbf{J},\mathbf{h}$ 
we can write the average log-likelihood more concretely as
\begin{eqnarray}
    \mathscr{L}_D(\mathbf{J},\mathbf{h})_{\boldsymbol{\lambda}} 
    &=& \sum_i h_i\left<s_i\right>_{e} + \sum_{ij}J_{ij}
    \left<s_is_j\right>_{e}+
    \mathcal{F}(\mathbf{J},\mathbf{h})  -
     \sum_{i} \lambda_{i} |h_i|^2 - \sum_{ij} \lambda_{ij} |J_{ij}|^2
     \label{eq:regularized-log-likelihood}
\end{eqnarray}
where $e$ stands for empirical expectation value,
\textit{i.e.} $\left<s_i\right>_{e}=\frac{1}{M}\sum_{m=1}^M s_i^{(m)}$, etc.
Since only empirical expectation values 
enter in the function to be maximized
it follows that the optimal parameter values 
are determined only by them.
This is true both with penalties and without.
One says that these empirical expectation values are
\textit{sufficient statistics}.

Let us now consider the case without penalties.
Maximizing the likelihood \eqref{eq:regularized-log-likelihood}
with respect to $\mathbf{J}$ and $\mathbf{h}$ for given samples $D=\{\mathbf{s}^{(m)}\}_{m\le M}$
is then the same as the minimization of the right-hand side of \eqref{eELT}, but where the 
values of ensemble averages (the parameters in \eqref{eELT})
take now the values of the sample averages (the parameters in  \eqref{eq:regularized-log-likelihood}).
We state this important fact as

\vspace{5pt}
\noindent \textbf{Max-entropy principle}: \emph{
Maximum-likelihood inference of parameters in an exponential 
family is equivalent to maximizing 
the entropy of a probability distribution given empirical expectation
values of the functions of the variables multiplying the parameters.
}
\vspace{5pt}

A large literature starting from Jaynes assigns an independent
epistemological importance to the max-entropy principle
\cite{Jaynes,Jaynes2,jaynes3}.
We rather subscribe to the alternative point of view 
that the undisputed practical utility 
of max-entropy is most easily explained by the importance 
of the maximum-likelihood inference criterion
and the ubiquity of probability distributions 
in exponential families.
In statistics this side has been argued
in the framework of \textit{information geometry}~\cite{Amari85,Amari21}.
An argument formulated in the language of DCA 
and applications to biological data analysis can be found in~\cite{Aurell}.
From the point of view of thermodynamics and statistical mechanics
the most commonly held point of view is indeed that Gibbs-Boltzmann 
distributions have objective existence, and are not only expressions of
ignorance and/or limits on what empirical averages can be determined.
For a discussion with references to the literature on the foundations of statistical mechanics, see \cite{auletta2017}.

\subsection{The two main DCA methods: nMF and PLM}
\label{sec21.2.3}
Naive mean-field inference (nMF), or simply 
mean-field inference, is a versatile and computationally
efficient DCA procedure which can be derived in several
different ways.
The most straight-forward is to say that it treats probability 
distributions of the Ising/Potts type as if they were Gaussian
probability distributions over continuous variables.
The matrix of the quadratic form of the Gaussian is the inverse
correlation matrix. The naive mean-field inference procedure 
applied to Ising variables is hence
\begin{equation}\label{eMFC}
    \mathbf{J}_*^{nMF} = -\boldsymbol{\chi}^{-1},
\end{equation}
where $\boldsymbol{\chi}$ is the empirical correlation matrix
and $*$ indicates inferred value.
When the problem is under-sampled the empirical correlation matrix
does not have full rank, and the inverse does not exist.
\textit{Regularized} nMF can hence be seen as regularized matrix inverse.
$L_2$-regularized nMF is explicitly so, in that it can be written as
as well-defined (and simple) matrix transformation \cite{auletta2017}.
$L_1$-regularized nMF was used in the DCA context
in \cite{Jones-2012a}, and can be computed by convex optimization
techniques. 
The breakthrough paper for DCA in biological applications 
\cite{Morcos-2011a} used a regularization with pseudo-counts
which is equivalent to adding a matrix proportional to the identity
to the empirical correlation matrix before taking the inverse.
The naive mean-field inference formula \eqref{eMFC}
was first obtained in \cite{Kappen-1998a}. Other ways to obtain it
can be found therein and in the standard methodological reference \cite{ChauNguyen2017}.

Pseudo-Likelihood Maximization (PLM)
\cite{Besag-1975a,Ravikumar-2010a} is 
computationally more demanding than nMF inference but,
however, it has the advantage that it is statistically consistent. It has also proven to be a more
accurate predictor in many applications.
PLM is an attempt to match parameters not to 
the full probability distribution but only to the set of
conditional probabilities of each variable conditioned on all the others.
In the context of exponential models, optimization of the PseudoLikelihood function
is particularly simple. The conditioned variables do not intervene in the normalization constant and the parameters related to these variables cancel out.
%The reason that this is a simplification for models in exponential 
%models is that all variables and parameters that pertain to the 
%condition side then cancel out. 
Indeed, for Ising models the conditional probability of $s_i$ under the observation of the other variables $\mathbf{s}_{\backslash i}$ is
\begin{equation}\label{ePLME}
    p(s_i\mid\mathbf{s}_{\backslash i})  =\frac{1}{2} \Big[ 1+s_i\tanh\Big( h_i + \sum_{j:j\ne i} J_{ij} s_j \Big)\Big]\ 
\end{equation}
which only depends on the field $h_i$ and on the couplings $\{J_{ij}\}_{j:j\ne i}$. 
From this one can form the log-pseudo-likelihood per sample,
\begin{align} 
    &\mathscr{L}^i_D(\{J_{ij}\}_{j:j\neq i},h_i)  = \frac{1}{M}\sum_m \log\frac{1}{2}\Big[ 1+s_i^{(m)}\tanh\Big( h_i + \sum_{j:j\ne i} J_{ij} s_j^{(m)} \Big)\Big]\label{eq:pseudo-log-likehood}
\end{align}
which can be optimized over its $N$ parameters.
The log-pseudo-likelihood can be regularized 
in the same manner as log-likelihoods. The authors of \cite{Ravikumar-2010a},
an article where the focus was on recovering the sparsity structure,
used an $L_1$ penalty. In most later applications of DCA 
instead an $L_2$ penalty was used, following \cite{Ekeberg-2013a,Ekeberg-2014a}.

It follows from above that the outcome of PLM is
a set of values $h_{*,i}^{PLM}$, one for each $i$, and a set 
$J_{*,ij}^{PLM,i},J_{*,ij}^{PLM,j}$, two for each pair $(ij)$.
The underlying probability distribution on the other hand
only contains one parameter $J_{ij}$ for each pair $(ij)$.
As part of PLM one therefore needs an output precedure.
One possibility is to sum all the log-pseudo-likelihood
functions 
\eqref{eq:pseudo-log-likehood}
and maximize all parameters at the same time.
This substitutes $N$ optimization problem on $N$
parameters with one optimization problem
on $N(N+1)/2$ parameters. Typically this increases
the computational burden.
An alternative procedure is to take for
the final inferred parameters the averages of the separately
inferred values:
\begin{equation}
J_{*,ij}^{PLM} = (J_{*,ij}^{PLM,i}+J_{*,ij}^{PLM,j})/2.
\end{equation}
In practice the second procedure is more used because it is computationally easier, and the first has not led to more accurate predictions.

\subsection{Advanced DCA methods}
\label{sec21.2.4}
Naive mean-field inference is built on a mean-field approximation
of the thermodynamic potentials.
This can be generalized by improving the approximation
which has been done by adding an 
Onsager reaction term.
The resulting inference procedure is called
Thouless-Anderson-Palmer (TAP) \cite{TAP}
and substitutes the equality in
\eqref{eMFC} with a quadratic transformation 
\cite{Kappen-1998a}. 
TAP inference has also been derived 
%as the first term beyond naive mean-field in a systematic expansion (Plevka expansion) in interaction strength \cite{}
by methods of information geometry \cite{Tanaka2000,Amari00informationgeometry}.

A second type of generalization is obtained by starting from more 
complicated trial functions.
Considering all pair-wise trial functions together with correlation-response
leads to an iterative procedure called \textit{susceptibility propagation}
\cite{MezardMora}. Later it was shown that the fixed point of susceptibility propagation can be computed in closed form \cite{Nguyen2012,Ricci_Tersenghi-2012}.
This gives an inference formula of  the same general form as \eqref{eMFC}
involving trancendental functions but still factorized over parameters 
to be retrieved.
For appropriately chosen models all the above methods outperform naive mean-field inference. Still, these methods are limited when the samples are drawn from ``clusterized'' dataset, such as low-temperature samples from the Ising model~\cite{nguyen2012mean,decelle2016solving}. For a unified treatment, see \cite{ChauNguyen2017}.

A major advance in the analysis of PLM 
and similar DCA algorithms was obtained in 
\cite{Vuffray-2016a,Bachschmid-Romano2017,Berg-2017a,Lokhov-2018a}
building on earlier results in 
\cite{Ravikumar-2010a} and \cite{Bresler-2015a}.
In all these papers were proven results of the type
that perfect structure recovery is achievable with only 
$M \sim \log N$ samples when the interaction graph is sparse, that there
is a gap between the smallest nonzero coupling and the set
of strictly zero couplings, and which successively weakened additional assumptions.
In \cite{Vuffray-2016a} 
it was shown for the Regularized Interaction Screening Estimator (RISE) 
algorithm that if the goal is to recover top-$k$ predictions
a gap is not needed, and in  \cite{Lokhov-2018a} a similar result was shown for
PLM (in section S1 of Supplementary Material).
The authors of  \cite{Vuffray-2016a} showed that RISE is asymptotically
optimal (in a certain sense)
and rigorously better than PLM (under certain assumptions).
In \cite{Berg-2017a} a further level of optimality was considered when 
some summary characteristics of the distributions over parameters 
was assumed known, and the optimum was computed using the replica method.
At this time, all these methodological advances are 
however limited 
to bounded maximum interaction strength 
and bounded degree of the interaction graph.

\subsection{DCA in computational biology}
\label{sec21.2.5}
The origins of the interest in DCA for biological data analysis 
can be traced to 
maximum-entropy arguments 
advanced in \cite{Schneidman-2006a}. 
In the terminology of above, that paper proposed 
naive mean-field inference of an Ising model.
%That means, as predictions of interactions they considered 
%elements of the inverse correlation matrix, and not
%elements of the correlation matrix.
Four years earlier, in a paper never published in a major 
journal, and therefore for a long time not widely known,
it was proposed in \cite{Lapedes2012}
that the parameters of a Potts model 
inferred from protein sequences in a protein family 
are good predictors of physical proximity.
Such amino acid-amino acid spatial contacts (residue-residue contacts)
contain important information on protein structure, non-local 
in the sequence. It can therefore be used (and was later used) as part of a protein structure prediction pipeline.
Residue-residue contacts also make sense as an evolutionary mechanism 
which contributes to the total biological fitness of the organism.
The technical term for such non-additive contributions to biological fitness
is \textit{epistasis}.
The relation between epistasis and DCA was reviewed in
\cite{Gao} and \cite{ZA-2}.

The flavour of DCA used in 
\cite{Lapedes2012} was maximum likelihood, computed by 
an iterative method. 
%It was therefore computationally quite expensive and the results were not as outstanding as obtained later. The principal reason, as understood later, was  that DCA needs more protein sequences in a protein family than typically available at that time.
The same problem was later addressed
for a well-characterized family of 
bacterial signal transduction pathways \cite{Weigt2009}.
Since for that data it proved possible to predict 
physical contacts between enzymes in the pathway which 
could be validated from other points of view,
this was an important step forward toward biological relevance.
The DCA implemented in \cite{Weigt2009}
was however susceptibility propagation
\cite{MezardMora} and thus also fairly 
computationally costly.
At the same time
residue-residue contacts were
derived from tables of sequences 
by another method not explicitly
in the DCA family \cite{Burger-2010a}.

The first result using DCA to predict
residue-residue contacts in proteins and which had wide resonance
used naive mean-field inference 
\cite{Morcos-2011a}, with a regularization using pseudo-counts.
DCA built on mean-field inference 
with other regularization schemes were introduced in the 
same application area in
\cite{Hopf-2012a,Jones-2012a,Andreatta-2014a}.
The second main type of DCA in computational biology 
has been pseudo-likelihood maximization
\cite{Ekeberg-2013a,Ekeberg-2014a,Feinauer-2014a},
and has been considered somewhat 
superior to naive mean-field inference in this
application domain \cite{Cocco-2018a}.
Later versions of DCA on the protein structure
prediction problem were 
meta-algorithms incorporating also other
information sources
\cite{Jones-2015a,Golkov-2016a,Hopf-2017a,Ovchinnikov-2017a}; although exhibiting higher performance
they (and the earlier undiluted DCA methods) have more recently
been overtaken by AI/deep learning methods~\cite{Senior-2020,Hiranuma-2021}. For other biological inference tasks with less abundant training data and/or where the goal is to uncover new biology, DCA (usually either mean-field of pseudolikelihood) remains an important tool \emph{cf.} \cite{Baldassi-2014a,Figliuzzi-2016a,Uguzzoni-2017a,DeLeonardis-2016a,Skwark-2017a,Schubert2018,Zeng-COVID19}. 

To end on a forward perspective, we consider human genome data.
The human genome contains about $3\cdot 10^9$ nucleotides out of which about
$60\cdot 10^6$ are in exons (parts of genes that code for proteins).
About a decade ago several large meta-studies on the origin of human obesity were published in leading journals where $M$ (total number of patients) was on the order of hundred of thousands or millions \cite{Speliotes2012,Bradfield2012,Locke2015}. 
Those studies were mainly built on exon sequencing, and with some filtering
out of amino acids fixed or almost fixed in the whole human population,
the effective number of loci of variability ($N$) was on the order of millions. 
The under-sampling ratio ($\sim M/N$) was hence in these studies
from order one to order $10^{-2}$. We formulate the challenges which 
pose themselves as

\vspace{5pt}
\noindent \textbf{DCA, human-scale genomic data}: \emph{
To design smart speed-ups yielding interesting 
predictions and are computationally feasible on the human exome scale
($N\sim 10^6$) and eventually on the human genome scale ($N\sim 10^9$).}
\vspace{5pt}

{We note that determining the
largest elements or set of largest elements of large correlation matrices
is known in computer science as 
``the light bulb problem''. The challenge can hence be considered 
an analogous problem from the perspective of DCA.
While the light bulb problem is not computationally hard 
(the elementary algorithm scales as $N^2$)
it is still not trivial in practice when $N$ is in the range of millions or billions.
Advanced algorithms are known to work 
in sub-quadratic time, but with significant pre-factors,
and only appear to become competitive for $N$ as large as $10^5$~\cite{Valiant2015}.}

\subsection{From DCA to Restricted Boltzmann Machines}
\label{sec21.2.6}
While all the previous approaches to DCA are concerned with the inference of pairwise interactions of an underlying statistical model, it is quite natural to consider generalizations that can also adjust for higher moments of the distribution. The most (seemingly) natural way would be to add higher-order terms in the Hamiltonian. However, this seems not reasonable since in order to include all possible terms of $n$-body interactions, the number of parameters to add is of order $\mathcal{O}(n^N)$. To avoid this issue, it is possible to use hidden-variable models. In these models, it is considered that a subset of nodes are not observed in the dataset but can still help in the modelization of the statistical properties when integrating them out. When integrated, they create effective many-body interactions between the variables to which they are connected. This type of approach has given rise to the celebrated Restricted Boltzmann Machine in the field of Machine Learning~\cite{smolensky1986information,hinton2002training}.

\vspace{5pt}
\noindent \textbf{RBM, definition and learning}: \emph{
To design a practical machine, the Hamiltonian of the model is defined on a bipartite graph. A first layer regroup the variables that are observed (from the dataset), named visible nodes, while the other layer is made of hidden nodes whose role is to induce effective interactions between those in the first layer.}
\vspace{5pt}

In its formulation, we usually distinguish the notation between the visible, $s_i$, and the hidden nodes $\tau_a$, yet both of them takes value in $\pm 1$. This led to the following Hamiltonian:
\begin{equation}
    H_{\mathbf{w},\mathbf{h}}(\mathbf{s},\boldsymbol{\tau}) = -\sum_{ia} w_{ia} s_i \tau_a - \sum_i h_i s_i - \sum_a h_a \tau_a. \label{eq:rbm}
\end{equation}
The bipartite structure, the fact that there is no coupling between two visible nodes or two hidden nodes, is a crucial aspect here to make Monte Carlo sampling much easier than in a fully connected model, making it practical to use the maxium likelihood approach in order to learn the parameters $\mathbf{w},\mathbf{h}$. 
As  one can see from \eqref{eq:rbm}, the distribution over the set of visible nodes, after integrating the hidden nodes, is no longer an exponential distribution and in particular, can adjust for higher moments of the dataset. This model has been used in the context of proteins~\cite{shimagaki2019selection,bravi2021rbm,tubiana2019learning}, but has not shown a further progress in the contact prediction task, and is thus more dedicated to extract useful features from a given dataset.

%\section{SBM aka Stochastic Block Model and community detection}\label{sec21.3}
%\section{Community detection and the Stochastic Block Model: the art of inferring clusters in a spin glass world}\
\section{Community detection: the art of inferring clusters in a spin glass world}
\label{sec21.3}

In the COMDEC problem, rather than inferring the parameters of the model like in DCA, the goal is the following: given a graph and an hypothesis on the nature of the interactions between its nodes, one has to infer the nodes states (\textit{i.e.} the ``community'' to which each node belongs to). The two standard hypotheses on the interactions are the \emph{assortative} case, meaning that the graph has been generated in a way that connections are statistically more present between nodes of the same community; this, in turns, make this problem a cousin of the Ising model (or Potts model when there are more than two hidden communities) with ferromagnetic interactions on a random graph. It is relevant to model, \textit{e.g.} friendships on social networks where people with similar opinions tend to be more friends. In the \emph{disassortative} case it is the opposite: links between members of different communities are favored, relating this model to an anti-ferromagnetic spin model. Ecological networks of predators-preys are of this nature.

To deal with this kind of problem, a successful approach has been to design a generative model that, given certain parameters, generates a graph where communities have been planted by construction. The paradigmatic generative model for graphs encoding community structures is the Stochastic Block Model (SBM) \cite{HOLLAND1983109,abbe2017community,moore2017computer}. With the SBM and its parameters in hand, it is then possible to design an inference procedure for the community structure, with the parameters being known. But also a learning algorithm which intend to estimate the values of the parameters given a particular family of models. 

\vspace{5pt}
\noindent \textbf{The Stochastic Block Model}: \emph{The SBM is a generative model of graphs with communities. As is often the case in Bayesian inference, first a generative model is defined before working out the form of the posterior distribution needed in order to estimate the parameters of the model.}
\vspace{5pt}

Many heuristic approaches have been considered for COMDEC~\cite{fortunato2010community}. For instance simple spectral methods were developed in order to partition a graph using the eigenvectors of variations of the adjacency matrix and/or Laplacian of the graph; later we are going to present state-of-the-art spectral clustering algorithms. But the lack of a proper model for how the graph communities were generated prevented to do much better than these simple approaches. The SBM was designed for that purpose. It is rooted on random graphs with additional ingredients forcing the network to develop a partition in the generation process. The ingredients are: 
\begin{itemize}
    \item the probabilities $\{n_a\}_{a\le q}$ of a node to belong to the community $a\in \{1,\ldots,q\}$;
    \item the probability matrix $\{p_{ab}\}_{a,b\le q}$ to have a link between two randomly chosen nodes, one in community $a$ and the other in $b$;
    \item a hidden (or ``planted'') partition of the nodes into $q$ communities or groups: $\{t_i\}_{i\le N} \in \{1,\dots,q\}^N$.
\end{itemize}

Major differences arise when considering different scaling regimes. In the \emph{dense regime} of the SBM, meaning $p_{ab}=\mathcal{O}(1)$ when $N\to+\infty$, the generated graphs have many ($\mathcal{O}(N^2)$) links, each node being connected to $\mathcal{O}(N)$ other ones. The \emph{logarithmic degree regime} corresponds to $p_{ab}=\mathcal{O}(\ln(N)/N)$ \cite{abbe2015exact}. Finally, in the \emph{sparse regime} $p_{ab}=c_{ab}/N=\mathcal{O}(1/N)$ the generated graphs have $\mathcal{O}(N)$ links and most nodes are connected to a finite number of other ones in the thermodynamic limit, up to few ``hubs''. In the dense regime, the communities can be ``read off'' from the degrees of the nodes, which makes the inference task rather straightforward using standard spectral methods. The real challenge arises in the sparse case, where completely novel ideas are needed, from a conceptual and algorithmic point of views, and where techniques from spin glasses have been particularly fruitful. We will thus focus on this sparse regime.

We now expose the four main algorithmic approaches which have altogether lead to a leap forward in the understanding of the COMDEC problem. We start with the Bayesian inference approach, given that it is in this context that one of the most striking phenomenon in COMDEC/SBM has been discovered: below the \emph{detectability transition}, it is information-theoretically impossible to distinguish a graph generated from the SBM from a purely random one.

\subsection{Approach 1: Bayesian inference}
Given the parameters of the SBM and under the assumption that the nodes partition and the observed graph $G$ (parametrized by its edges $A_{ij}$ between nodes $i$ and $j$) were drawn according to this generative model, it is possible to write down the joint law of $G$ and the partition:
\begin{align}
    P(G=\{A_{ij}\},\{t_i\}\mid \{n_a\},\{p_{ab }\})  &= P(G\mid \{t_i\},\{n_a\},\{p_{ab }\})P(\{t_i\}\mid \{n_a\}) \nonumber \\
     &= \prod_{i < j} p_{t_i t_j}^{A_{ij}} \left(1-p_{t_i t_j} \right)^{1-A_{ij}} \prod_i n_{t_i} . \label{eq:def-sbm}
\end{align}
Hence, the probability to put a link between two nodes now depends on the community of each node and the probability to have a link precisely between those two nodes.

\vspace{5pt}
\noindent \textbf{Inference of the groups}: \emph{Having defined our generative model, we need to decide how to infer the probability that a given node of the graph belongs to one of the possible groups.}
\vspace{5pt}

With this generative model it is now theoretically possible to infer the nodes states by computing their marginal probabilities to belong to one of the communities. This can be done using Bayes theorem. Given parameters $\{n_a\}$ and $\{p_{ab}\}$ (in addition to $G$) the posterior reads
\begin{equation}
    P(\{t_i\} \mid G,\{n_a\},\{p_{ab }\}) = \frac{P(G,\{t_i\}\mid  \{n_a\},\{p_{ab }\})}{\sum_{\{t_i\}} P(G,\{t_i\} \mid  \{n_a\},\{p_{ab }\})}. \label{eq:def-ham}
\end{equation}
It relates the probability of an assignment to the generative model that we defined in  \eqref{eq:def-sbm}. This distribution can be rephrased as the Gibbs-Boltzmann distribution of a Potts model by taking (minus) the log of the un-normalized probability \eqref{eq:def-ham} (recall that $p_{ab}=c_{ab}/N$):
\begin{align}
    %\mathcal{H}(\{t_i\}) = \sum_i \log(n_{t_i}) + \sum_{(ij) \in \mathcal{E}} \log\left( \frac{p_{t_i t_j}}{1-p_{t_i t_j}} \right) - \sum_{(ij) \notin \mathcal{E}} \log\left( 1-p_{t_i t_j} \right)
        \mathcal{H}(\{t_i\}) &=-\sum_i \log n_{t_i} - \sum_{(ij) \in \mathcal{E}} \log p_{t_i t_j} - \sum_{(ij) \notin \mathcal{E}} \log( 1-p_{t_i t_j} )\nonumber \\
        &=-\sum_i \log n_{t_i} - \sum_{(ij) \in \mathcal{E}} \log c_{t_i t_j} + \frac1N\sum_{(ij)\notin \mathcal{E} } c_{t_i t_j}  +C
\end{align} 
(where the second equality is true up to a negligible $o(N)$ correction) with $\mathcal{E}=\{(ij):A_{ij}=1\}$ is the set of edges of $G$, and $C$ is an irrelevant constant. It is possible to write the first term as $-\sum_{i,a} \delta_{a,t_i} \log n_a$, and the second one too using a similar trick with Potts interacting terms $\delta_{a,t_i}$ and $\delta_{b,t_j}$ living on the interaction graph $G$. The third term is interesting because it describes an interaction between nodes that do \emph{not} share an edge, which is rather uncommon in typical disordered spin systems. 

\vspace{5pt}
\noindent \textbf{Both edges and non-edges contain information}: \emph{A specific feature of the SBM is  that not only an observed edge of $G$ yields information, but also the absence of an edge does so. Each edge results in a $\mathcal{O}(1)$ interaction in the equivalent Potts model, which, \textit{e.g.} is ferromagnetic in the assortative case; absence of an edge yields a ``weak'' $\mathcal{O}(1/N)$ interaction term, which overall compete with the ``edge'' interactions due to their large number.}
\vspace{5pt}

Hence, the inference task of computing the marginals in the SBM is equivalent to the estimation of the local mean magnetizations of a peculiar Potts model, with a field generated by the non-edges which prevents the Potts variables to end up in a ferromagnetic state where most have same value. Said differently, this field enforces phase separation, \textit{i.e.} appearance of separated communities. The standard approach for sampling would be to use Monte Carlo methods, but, as we will see, Belief Propagation (or equivalently the cavity method) can be used in a very efficient way.

The second task is to estimate, or learn, the parameters of the model. Again using the Bayes theorem, we can write a posterior probability distribution but this time on the parameters:
\begin{equation}
    P(\{n_a\},\{p_{ab}\}\mid G) = \frac{P(G\mid \{n_a\},\{p_{ab}\}) P(\{n_a\},\{p_{ab}\}) }{P(G)}. \label{eq:like-param}
\end{equation}
Without being too specific, the simplest case to deal with is by considering the absence of prior on the parameters, and to look only at a (possibly local) maximum of the probability distribution by performing a gradient ascent. The above expression therefore tells us that we need to maximize the free energy  $\ln P(\{n_a\},\{p_{ab}\}\mid G)$ of our SBM with respect to the parameters. Interestingly, this gradient can be expressed in terms of the mean values computed during the inference problem, and using the expectation-maximization method we get the following iterative algorithm:
\begin{align}
    p_{ab}^{t+1} = \frac{1}{n_a n_b} \sum_{(ij) \in \mathcal{E}} \langle \delta_{t_i,a} \delta_{t_j,b} \rangle_t , \qquad n_a^{t+1} = \frac1N{\sum_i \langle \delta_{t_i,a} \rangle_t}, \label{eq:sbm:learn1} 
\end{align}
where $\langle \cdot \rangle_t$ means an average taken using the parameters at iteration $t$.

\vspace{5pt}
\noindent \textbf{Learning the parameters of the models}: \emph{In the context of the RBM, it is also possible using the Bayes theorem to define a leaning algorithm to estimate the graph's parameters if unknown.}
\vspace{5pt}

\subsection{Phase diagram and the detectability transition}\label{phasediagCOMDEC}

%Before discussing the learning properties of this model, it is interesting to understand the phase diagram of the model. 
The inference performance in the SBM have received a particular attention in the symmetric case, where the intra-community and inter-community connectivity is the same for all groups: $c_{aa}=c_{\rm in}$, $c_{ab}=c_{\rm out}$. Tuning the difference between these values allows to interpolate between graphs where no communities are present (by imposing that an edge is present between two nodes independently of the node's group) and thus inference is impossible, to a regime where the graph is made of disconnected clusters and inference is trivial. It was shown in  \cite{decelle2011inference} that along this interpolation, the model exhibits a second order phase transition between a paramagnetic phase, where no information can be retrieved on the communities (in the large size limit) to a condensed phase where the equilibrium properties of the Boltzmann distribution are dominated by a state with a strong overlap with the planted community: this is the \emph{detectability transition} which has revived the research on COMDEC. It was later on rigorously validated \cite{massoulie2014community,mossel2015reconstruction,mossel2018proof} (including in more general settings, see \cite{abbe2017community}). 

\vspace{5pt}
\noindent \textbf{Detectability threshold in the two-groups symmetric SBM}: \emph{For a graph drawn from the SBM, non-trivial inference of the communities is possible in the large size limit if and only if $(c_{\rm in} - c_{\rm out})^2>2(c_{\rm in} + c_{\rm out})$. If this condition is not met, it is impossible to distinguish a graph from the SBM from a purely random one.}
\vspace{5pt}

In both the sparse and dense regimes when the inference occurs on the ``Nishimori line'' of the Hamiltonian \cite{nishimori2001statistical}, \textit{i.e.} is done in the Bayesian optimal regime where the SBM parameters are known, it implies that no spin glass phase can be present as strong concentration-of-measure effects take place; this is true more generically for optimal Bayesian inference problems \cite{barbierPanchenko}. Therefore the replica symmetric solution for the free energy is exact \cite{10.1093/imaiai/iaw017,coja2018information,barbier2019mutual,abbe2021stochastic}. 

The phase diagram of the model can be obtained by investigating the local stability of the Belief Propagation (BP) equations close to the paramagnetic fixed point. The BP equations for the SBM are given by
\begin{equation}
    \psi^{i \rightarrow j}(t_i) \propto n_{t_i} \prod_{k  \neq j,i} \sum_{t_k} p_{t_i t_k}^{A_{ik}} \left(1-p_{t_i t_k} \right)^{1-A_{ik}} \psi^{k \rightarrow i}(t_k)
\end{equation}
where $\psi^{i \rightarrow j}$ is the marginal of node $i$ when the link $(ij)$ has been removed. We note again the interesting property of the SBM that ``non-edges'' yield a contribution in the BP equation. In the symmetric case discussed in the previous paragraph, we can therefore easily identify the paramagnetic fixed point as the one where all the BP messages are uniform $\psi^{i \rightarrow j}(t_i) = 1/q$. The perturbation near this solution can be used to find the parameters at which the condensed phase starts, \textit{i.e.} when the paramagnetic fixed point becomes unstable. It is to be noted that this is the analog to the 
de Almeida-Thouless line for spin glasses~\cite{Almeida_1978}.

This description is not always correctly describing the physics of the problem. In fact, the presence of a first order phase transition can complicate the phase diagram. It is possible to verify if it is the case by analyzing the metastable states nearby the planted community structure. In~\cite{decelle2011asymptotic}, it is shown that when dealing with some graph's parameters (higher connectivity and number of communities), the SBM can have this type of dynamical transitions preventing efficient algorithms to infer the groups, yielding to even richer phase diagrams with, standing in between the detectability transition and the dynamical one, the presence of an algorithmically ``hard phase'' (also called statistical-to-computational gap). In this phase, the communities do have a statistical reality, yet the inference process is plagued by the ``paramagnetic'' solution to which the BP algorithm converges in absence of good initial condition provided by an oracle. To distinguish clearly which of the two states (possible recovery or not) dominates the equilibrium distribution and hence to understand if the planted solution is statistically distinguishable from the paramagnetic one, it is enough to compare the free energy of both states. We end up with a picture where the physics of phase-coexistence explains here the different phases of the inference process in the SBM. This past decade it has been understood that the same type of phase diagram and phenomenology holds more generically in a broad class of high-dimensional Bayesian inference and learning problems such as, \textit{e.g.} in the generalized linear model \cite{barbier2019optimal}, see also \cite{RSB}.

\vspace{5pt}
\noindent \textbf{Statistical-to-computational gap}: \emph{For certain parameters of the SBM with more than two communities, a dynamical transition may prevent efficient algorithms to saturate the detectability transition: an algorithmically hard phase emerges. This phenomenon occurs in many other high-dimensional inference problems.}
\vspace{5pt}

%\subsection{Learning instabilities}
Finally, the learning equations \eqref{eq:sbm:learn1} can also be analyzed in some restricted cases~\cite{kawamoto2018algorithmic}. Interestingly, for a given type of graphs parametrized by the degree structure and number of communities, it is possible to understand whether the EM equations can learn the true graph's parameters. Again, the typical phase diagram is made of distinct regions. First, an uninformative region, where basically the learning dynamics is stuck. This region encompasses both the regime where the created graph has no statistically relevant community structure and also a regime where the communities are present but the dynamics is not able to drive the parameters toward their correct values. Finally, in the last region, the EM equations might bring the meta-parameters to their correct values. 

\subsection{Approach 2: Modularity optimization}

While in the previous section COMDEC was settled by using first a generative model and second an inference algorithm to recover the graph's structure conditionally on this model, other approaches have studied settings where the model mismatched the dataset, which is  particularly relevant when dealing with real data for which the generative process is unknown. One of these approaches~\cite{PhysRevE.67.026126,newman2006modularity} is based on the maximization of the \textit{modularity} of the graph. Given a graph and the degree $k_i$ of each nodes, the modularity for partitioning two clusters is given by
\begin{equation}
    Q(\bm{s} \mid G, \mathbf{k}) = \frac{1}{4M}\sum_{i < j} \Big( A_{ij} - \frac{k_i k_j}{2M} \Big)s_i s_j,
\end{equation}
$M$ being the total number of edge and the variables $s_i$ represents the group assignment (here $\pm 1$) of a given node. The modularity tends to optimize the number of edges within a community with respect to the expected number of edges resulting from a random graph. In the first approaches, the modularity was optimized in order to find the best possible partition of the nodes. Yet, since the problem is hard, it is probably impossible to design a polynomial algorithm that finds in general the optimal value of $Q$. This problem was then addressed in two successive works~\cite{zhang2014scalable,schulke2015multiple} where a temperature is introduced in order to define the Gibbs-Boltzmann distribution associated to the modularity:
\begin{equation}
    P(\bm{s}\mid G, \mathbf{k}) = \frac{\exp\left( - M\beta Q(\bm{s}\mid G, \mathbf{k}) \right)}{\mathcal{Z}( G, \mathbf{k})} .
\end{equation}
The effect of the temperature is to help avoiding possible overfitting of the modularity, since the marginals obtained by this mean have now to take into account the entropy of the inferred clusters. With this approach it is now possible to study the phase diagram of the model in temperature and as a function of the true graph parameters when dealing with networks generated by the SBM for instance. Interestingly, a spin glass phase is often found at low enough temperature, implying that the optimization problem becomes very hard. At the same time, it emerges a retrieval phase where the communities can be retrieved easily, and which underlines the importance of correctly selecting the temperature.

\subsection{Approach 3: Spectral clustering}

A particularly fruitful approach to COMDEC is spectral clustering. Spectral algorithms are very practical, given that they use only basic linear algebra (eigen-decomposition and singular value decomposition) to partition the graph. All the point is therefore to design smartly the matrices to analyze. Variations of the adjacency matrix $\mathbf{A}$ of the graph or of its graph Laplacian $\mathbf{L}=\mathbf{D}-\mathbf{A}$ ($\mathbf{D}$ being the diagonal matrix of nodes degrees) are natural choices \cite{von2007tutorial}, but all suffering from the curse of eigenvector localization: their eigenvectors tend to encode local graph structures (like atypical nodes of high connectivity) rather than global ones such as potential communities. But much better choices exist. 

A breakthrough in COMDEC came from the discovery that by analyzing the so-called \emph{non-backtracking matrix} $\mathbf{B}$ introduced in the theory of zeta functions \cite{hashimoto1989zeta}, one could recover the communities in the SBM down to the detectability threshold whenever no hard phase is present, or down to the best known algorithmic threshold (set by BP), see \cite{doi:10.1073/pnas.1312486110}. This matrix indexed by the directed edges $i\to j\in \mathcal{E}_d$ of the graph ($\mathcal{E}_d$ being of size twice the one of the edge set $\mathcal{E}$) is defined as 
\begin{equation}
B_{i\to j,k\to \ell}=\delta_{jk}(1-\delta_{i\ell})    
\end{equation}
and therefore encodes all possible non-backtracking paths of length two of the graph. As such it is also called ``edge adjacency operator''. When the graph is sparse, its dimension is of the same order as the number of nodes. Considering again the two-groups symmetric SBM, let the average degree be $c=(c_{\rm in}+c_{\rm out})/2$. The key properties of this non-symmetric matrix $\mathbf{B}$ are: $(i)$ it is much less sensitive to the eigenvector localization problem (\textit{i.e.} to atypical high-degree nodes) than standard operators; $(ii)$ its complex eigenvalues are (asymptotically) confined in a the complex disk of radius $\sqrt{c}$ except for two real eigenvalues $\lambda_1\approx c$ and $\lambda_2\approx (c_{\rm in}-c_{\rm out})/2$; $(iii)$ the eigenvector $\boldsymbol{v}$ associated with $\lambda_2$ is strongly correlated to the hidden communities. More precisely, letting $\hat t_i={\rm sign}\sum_{j :j\to i\in \mathcal{E}_d} v_{j\to i}$, then $\{\hat t_i\}_{i\le N}$ correlates with the hidden community structure almost as good as the BP estimate, itself conjectured optimal among polynomial algorithms and information-theoretically optimal when there is no statistical-computational gap (like in the two-groups symmetric SBM). 

Why the spectral algorithm associated to the non-backtracking matrix behaves so similarly to BP? Simply because they are closely related. We mentioned already in the Sect.~\ref{phasediagCOMDEC} that the phase diagram of the inference in the SBM could be studied through a linearization of the BP equations around the paramagnetic fixed point. But this idea can also be turned into a spectral algorithm as follows. Still considering the symmetric two-groups SBM, let the BP messages close to the paramagnetic solution be $\psi^{i \rightarrow j}(\pm1)=1/2\pm\delta^{i \rightarrow j}$. Thus, the vector $\boldsymbol{\delta}$ quantifies the first order deviation of the BP messages around the paramagnetic fixed point. Then the linearization of the BP equation can be re-written as 
\begin{align}
    \mathbf{B}\,\boldsymbol{\delta}=\Big(\frac{c_{\rm in}+c_{\rm out}}{c_{\rm in}-c_{\rm out}}\Big)\boldsymbol{\delta}.
\end{align}
Therefore the ``BP message deviation'' $\boldsymbol{\delta}$ is an eigenvector of the non-backtracking operator $\mathbf{B}$. It thus makes sense that this spectral approach matches closely BP, at least close to its paramagnetic transition. That its performance remains so good far from it is a surprising fact.

\vspace{5pt}
\noindent \textbf{Belief Propagation inspired spectral algorithms}: \emph{By defining a matrix related to the linearization of the BP equations around its paramagnetic non-informative fixed point, it is possible to saturate the BP performance using its spectral analysis, even without knowing the parameters of the model.}
\vspace{5pt}

These desirable properties extend when considering more than two communities: the estimator extracted from $\mathbf{B}$ remains competitive with the BP one until the BP algorithmic transition. Interestingly, the number of real eigenvalues escaping the disk in the complex plane even indicates the number of hidden communities while BP needs that prior information. So it seems that this spectral approach solves all problems at once: it is simply based on linear algebra, robust to localization and competitive with BP. But there are two catches. Firstly, even if the average degree of the graph is bounded the size of this (directed) edges-indexed matrix $\mathbf{B}$  can quickly become unpractical when it comes to diagonalize it. Moreover it is non-symmetric while numerical routines for linear algebra tend to be much more optimized for symmetric matrices. Secondly, despite being robust to localization, it turns out to be hyper sensitive to other type of ``noises'', \textit{i.e.} to deviations from the ideal setting where the graph is locally tree-like as in the sparse SBM. One example noticed, \textit{e.g.} in \cite{javanmard2016phase,ricci2016performance,zhang2016robust} is the appearance of very few small cliques (fully connected sub-graphs) in the graph $G$ which completely breaks apart the approach of \cite{doi:10.1073/pnas.1312486110} based on $\mathbf{B}$. See also \cite{moitra2016robust,feige2001heuristics} for robustness issues.

The first issue on the computational cost has been solved by introducing another linear operator $\mathbf{H}$ sharing the very same desirable properties of the non-backtracking operator while being symmetric and of smaller size $N\times N$: it is the \emph{Bethe Hessian} \cite{saade2014spectral} (recall $k_i$ are the nodes degrees) 
\begin{equation}
H_{ij}(r)=(r^2-1+k_i)\delta_{ij} -r A_{ij}.  
\end{equation}
It can be shown that any $r$ such that $\mathbf{H}$ possesses a vanishing eigenvalue corresponds to a real eigenvalue of the non-backtracking operator, thus its similar properties. Moreover, the minimum of its non-informative bulk of eigenvalues reaches $0$ when $r \approx \sqrt{c}$, so that the informative isolated eigenvalues whose associated eigenvectors correlate with the community structure are on the negative axis, a quite convenient property. The name of this matrix comes from the fact that it is proportional to the Hessian of the Bethe free energy (the matrix of second derivatives with respect to the spin magnetizations) of an Ising model living on the edges of the graph $G$, evaluated at the paramagnetic solution. The approach has also been extended to a close relative of the COMDEC problem, namely tensor principal components analysis. The authors of \cite{wein2019kikuchi} defined the \emph{Kikuchi Hessian} which, as the name suggests, is related to the Hessian of the Kikuchi free energy, a hierarchical free energy approximation whose first order is the tree approximation (\textit{i.e.} Bethe free energy) and that takes more and more local structures (such as loops) into account at the next levels of the hierarchy \cite{yedidia2003understanding}. But despite its many advantages, like the non-backtracking operator, the Bethe Hessian fails whenever the graph is not ``locally tree-like enough'', see the previous references.

An elegant proposal to cure this latter issue of lack of robustness of spectral approaches is the \emph{X-Laplacian} $\mathbf{L}_X=\mathbf{A}+\mathbf{X}$ \cite{zhang2016robust}. In this rather generic method to solve the eigenvector localization problem and improve noise robustness, a problem-dependent fine tuned Laplacian matrix is iteratively computed by regularizing the adjacency matrix $\mathbf{A}$ with a diagonal $\mathbf{X}$ strongly penalizing eigenvectors which are localized (roughly meaning that their norm is dominated by a few/sub-linear fraction of the components). Along the iterations of the learning procedure the most localized eigenvectors of $\mathbf{L}_X$ among those paired with the top eigenvalues living towards the end of the bulk, and which hide the informative eigenvalues/eigenvectors, are pushed back inside the bulk by the learned regularization $\mathbf{X}$. This has for effect to ``uncover'' the informative eigenvalues, left untouched by the procedure as their associated eigenvectors were delocalized in the first place.

\subsection{Approach 4: Convex relaxation}

Another success story is an algorithm rooted in computer science: convex relaxation and semi-definite programming \cite{boyd2004convex}. In \cite{javanmard2016phase} it is proposed to recover the communities in the symmetric two-groups SBM as follows. First, notice that an SBM instance can be mapped onto the \emph{$\mathbb{Z}_2$-synchronization problem} \cite{10.1093/imaiai/iaw017}: infer $\mathbf{x}\in\{-1,1\}^N$ given the gaussian corrupted observations 
\begin{align}
Y_{ij}\sim \mathcal{N}(m=\gamma N^{-1/2} x_ix_j,\sigma^2=1), \quad 1\le i<j\le N.\label{z2}    
\end{align}
An heuristic argument to see the connection with the SBM goes as follows: edges being present or not are, conditionally on the planted partition, independent Bernoulli random variables $A_{ij}\sim {\rm Ber}(c/N + x_i x_j(c_{\rm in}-c_{\rm out})/(2N))$. Their variance is, whenever $c_{\rm in}$ and $c_{\rm out}$ are close, approximately equal to $c/N$. Matching its first two moments with gaussian variables yields $\tilde Y_{ij}\sim \mathcal{N}(c/N + x_i x_j(c_{\rm in}-c_{\rm out})/(2N),c/N)$ which are equivalent in terms of information content to $Y_{ij}$ when setting the signal-to-noise ratio $\gamma=(c_{\rm in}-c_{\rm out})/(2\sqrt{c})$ in \eqref{z2}. This fruitful connection between the SBM and $\mathbb{Z}_2$-synchronization can be made rigorous and has been exploited to carry precise information-theoretic analyses \cite{korada2009exact,10.1093/imaiai/iaw017,dia2016mutual,lelarge2019fundamental}.

With this mapping in mind \cite{javanmard2016phase} considers the following convex relaxation of $\mathbb{Z}_2$-synchronization: given $\mathbf{Y}$,
\begin{equation}
    {\rm maximize} \ {\rm Tr} (\mathbf{X}\mathbf{Y}^\intercal) \ \mbox{s.t.} \  \mathbf{X}\succcurlyeq \boldsymbol{0} \ \mbox{and} \ X_{ii}=1 \ \mbox{for} \  i=1,\ldots,N. \label{SDP}
\end{equation}
Its $N\times N$ matrix solution is easily obtainable with a convex solver. The final estimator is then $\hat{\mathbf{x}}(\mathbf{Y})=\sqrt{N} T(\gamma) \boldsymbol{\nu}_1$ where $T(\gamma)$ is an appropriate scaling and $\boldsymbol{\nu}_1$ is the eigenvector of the non-negative definite solution of \eqref{SDP} with highest eigenvalue. Surprisingly, this estimator is \emph{almost} as good as BP; it does not quite saturate the detectability transition in the two-groups case (while BP does) but its algorithmic transition is extremely close to it. Moreover, by studying a certain vectorial spin model thanks to the replica and cavity methods, it is possible to precisely predict the asymptotic performance of this powerful procedure.

\vspace{5pt}
\noindent \textbf{$\mathbb{Z}_2$-synchronization/SBM equivalence}: \emph{The information-theoretic analysis of the stochastic block model is closely related to the analysis of the $\mathbb{Z}_2$-synchronization (or rank-one matrix factorization) problem. This mapping also serves as inspiration to design new inference algorithms.}

\section{Dynamic cavity method}\label{sec21.4}

In the previous sections the graphs represented static interactions. But what about inference from graphs encoding causality and/or time dependencies? In this section we present an approach exploiting message passing techniques. In the literature it is known as Dynamic Cavity. The natural way in which messages and cavity-like equations can be used to solve dynamical problems suggests a more general view of inference outside the realm of Gibbs-Boltzmann distributions which has been our focus this far.

\subsection{Definition, specificity and main problem}
\label{sec21.4.1}
The standard cavity method is a means to compute marginals of a Gibbs-Boltzmann distribution 
by exchanging messages~\cite{MezardMontanari}.
We consider a dynamics specified by an interaction graph of the same locally tree-like type
as where the cavity method has found its main applications. Let the history of variable $i$ up to time
$t$ be $X^t_i$, and
let the value of variable
$i$ at time $t$ be $\sigma_i(t)$. 
{For an Ising variable we formally define
$X^t_i=(\sigma_i(t_0),t_1,t_2,\ldots)$ 
where $\sigma_i(t_0)$ is the initial value of the spin
and $t_1,t_2,\ldots$ are the set of spin flip times.
For a categorical variable history one would additionally 
have to keep track of between which states the jumps happen.
}
The joint probability of all the variables 
is supposed to satisfy a high-dimensional differential equation 
(master equation) of the type
\begin{equation}
\label{eq:master-equation}
    \frac{d}{dt}P\left(\sigma_1,\ldots,\sigma_N,t\right) = \sum_{\sigma'_1,\ldots,\sigma'_N}
    \Gamma_{\mathbf{\sigma},\mathbf{\sigma}'} P\left(\sigma'_1,\ldots,\sigma'_N,t\right).
\end{equation}
The locally tree-like interaction graph describes the transition matrices
$\Gamma_{\mathbf{\sigma},\mathbf{\sigma}'}$ which
satisfy $\sum_{\mathbf{\sigma}}\Gamma_{\mathbf{\sigma},\mathbf{\sigma}'}=1$ for every
value of $\mathbf{\sigma}'$.
The joint probability over the histories of all the variables can then, up to technicalities,
be written 
\begin{equation}
\label{eq:transition-matrix-product}
    P^t(X^t_1,\ldots,X^t_N) = 
    \Gamma_{\mathbf{\sigma}(t),\mathbf{\sigma}(t-\epsilon)} \cdots \Gamma_{\mathbf{\sigma}(t_0+\epsilon),\mathbf{\sigma}(t_0)}
    \cdot P^0(\sigma_1,\ldots,\sigma_N,t_0).
\end{equation}
We either assume that the initial probability distribution $P^0$ is so far in the past that it does not matter, or that it only has the same dependencies as in
$\Gamma$. For instance, 
it can be factorized.
We now additionally assume that the probabilities 
of different variables to flip in a time interval $\Delta t=\epsilon$
are independent. This is natural in the continuous-time limit 
where these probabilities are given by 
$\Delta t\cdot r_i$ where $r_i$ is the instantaneous flip rate of spin $i$.
For the Ising ferromagnet with Glauber dynamics,
which we show as a numerical example in
Fig.~\ref{fig.iCME}, the rates are
\begin{align}
r_i(\sigma_i,\sigma_{\partial i})
= \alpha\frac{
e^{-\frac{J}{k_B T}\sum_{j\in\partial i}\sigma_i\sigma_j}}{
\sum_s e^{-\frac{J}{k_B T}\sum_{j\in\partial i}s\sigma_j}},
\end{align}
where $\alpha$ is a constant of dimension inverse time and 
$J$ is the pairwise interaction energy.
In this example
the dependencies in the joint probability distribution \eqref{eq:transition-matrix-product}
includes effects of the type that if $k$ and $j$ are both
in the neighborhood of $i$ as given by the energy function, 
they are also related by the denominator in the expression for the
rate $r_i$. The total statistical dependencies in \eqref{eq:transition-matrix-product} therefore include many local loops.
A systematic approach to resolve these loops is 
by graph expansion 
\cite{Altarelli2013,Gino2015}. This approach associates a pair of variables
$(X^t_i,X^t_j)$ to each link $(ij)$ in the original dependency
graph and imposes hard constraints $C_i$ that all variables of the type
$X^t_i$ in all links $(ij)$ take the same value.
The probability distribution \eqref{eq:transition-matrix-product}
can then equivalently be written 
\begin{equation}
\label{eq:transition-matrix-product-2}
    P^t\big(\{X^{t,(ij)}_i,X^{t,(ij)}_j\}\big) =
    \prod_i \Phi_i\big(X^{t}_i,\{X^{t,(ij)}_j\}_{j\in \partial i}\big)
    \prod_i C_i
\end{equation}
where the local loops have been resolved.
The first argument $X^{t}_i$ 
on the right hand of above can be any of the $X^{t,(ij)}_i$
as by the constraint $C_i$ they are all the same.
Using
the theory of Random Point Processes \cite{vanKampen}
the local weight functionals can be written
\begin{eqnarray}
\Phi_i\left(X^{t}_i,X^{t}_{\partial i}\right)
&=& \prod_{s=1}^{n} r_i\left(\sigma_i(t_s),\sigma_{\partial i}(t_s)\right)
\cdot e^{-\int_{t_0}^{t_1} r_i\left(\sigma_i(\tau),\sigma_{\partial i}(\tau)\right)
\, d\tau} \prod_{s=1}^{n} 
e^{-\int_{t_s}^{t_{s+1}} r_i\left(\sigma_i(\tau),\sigma_{\partial i}(\tau)\right) \, d\tau}
\end{eqnarray}
where we recall that 
$X^{t}_i$ is defined by $n$, 
the number of jumps of spin $i$ in a time interval $[t_0,t_f]$, the initial 
spin state, and the jump times.
For given $n$ the last time ($t_{n+1}$) in above is $t_f$. 

After applying the graph expansion
the right-hand side of \eqref{eq:transition-matrix-product-2} is like  
a Boltzmann weight with hard constraints in the standard cavity method. 
In the original formulation \eqref{eq:transition-matrix-product}
the marginal probability over one history is defined as
\begin{equation}
    P^t_i(X^t_i) = \sum_{\mathbf{X}_{\setminus i}} P^t(X^t_1,\ldots,X^t_N)
\end{equation}
and in the expanded graph we can first marginalize to the joint probability of the set $\{X^{t,(ij)}_i,X^{t,(ij)}_j\}_{j\in\partial i}$,
where all the $X^{t,(ij)}_i$ are the same due to the 
constraint $C_i$, and then marginalize separately over 
the $X^{t,(ij)}_j$.
The cavity (or Belief Propagation) output equation is
then
\begin{equation}
    P^t_i(X^t_i) = \sum_{X_{\partial i}^t} \Phi_i\left(X^t_i,X^t_{\partial i}\right)
    \, \prod_{j\in\partial i} \mu^t_{j\to (ji) }(X^t_j,X^t_i ).
       \label{eq:cavity-output-X}
\end{equation}
In above $\mu^t_{j\to (ji) }(X^t_j,X^t_i)$ (a message with two arguments)
follows from the graph expansion, 
and the egress node notation $(ji)$
indicates that these messages are 
actually passed around in the expanded graph.
The cavity (or BP) update equation is on the same level of abstraction
\begin{align}
    &\mu^t_{j\to (ji) }(X^t_j,X^t_i) =\sum_{X^t_{\partial j\setminus i}} \Phi_j\left(X^t_j,X^t_{\partial j}\right)  
    \prod_{k\in\partial j\setminus i} \mu_{k\to (kj)}(X^t_k,X^t_j).
    \label{eq:cavity-update-X}
\end{align}
The problem of using \eqref{eq:cavity-output-X} and \eqref{eq:cavity-update-X} is that the 
the argument is very high-dimensional. 
Various approximations have been introduced in the literature
to nevertheless make the dynamic cavity an efficient and accurate modelling approach.
As in this chapter we consider statistical inference we will not discuss the use
of dynamic cavity to retrodict the origin of epidemics and similar processes, for this
see \cite{Altarelli2014} and \cite{Lokhov2014}. To stay inside the sphere of physical problems
we will also not consider the recent use of dynamic cavity to model and predict the evolution 
of an epidemic \cite{Ortega2022,David2}.

\subsection{Dealing with the histories}
\label{sec21.4.2}
After deriving \eqref{eq:cavity-update-X}, the next step is to find a convenient parametrization for the histories. In a discrete time setting, a simple parametrization is to consider the values of the spins at different times. Each $X_i^t$ is then approximated by the values of the spins at different times, $X_i^t \approx (\sigma^t, \sigma^{t -\epsilon}, \cdots ,\sigma^0)$. To arrive at finite-dimensional messages one can then consider a closure on the last $n$ times, which means to take into account a memory of length $n\delta t$. This was the approach (for $n=2$) followed in \cite{Gino2015,Neri} when studying of the kinetic Ising model under synchronous update dynamics. A more advanced approach based on the matrix product expansion from quantum condensed matter theory was investigated in \cite{Barthel2018}.
%An alternative solution was proposed in \cite{Pelizzola,Gino2017} using a cluster variational method for a functional defined on the trajectories, one obtain equations that resemble \eqref{eq:cavity-update-X}, but where the graph of interaction takes into consideration the temporal correlation between the variables.

Continuous-time dynamics is problematic using both the above approaches.
In a series of papers reviewed in \cite{Dominguez2020} a continuous-time closure 
was introduced leading to a cavity master equation.
Apart from the kinetic Ising model (pair-wise interactions)
this versatile approach has also been applied 
with good results to the ferromagnetic $p$-spin model under Glauber dynamics \cite{pspindyn}, and to the dynamics of a focused search algorithm to solve the random 3-SAT problem in a random graph \cite{ksatdyn}. 
The method has also generalized to provide master equations for the probability densities of any group of connected variables \cite{David2}. 

Here we will exemplify by a recent development
closer in spirit to the form of the dynamic cavity embodied 
by \eqref{eq:cavity-output-X} and \eqref{eq:cavity-update-X}.
The fundamental object of the cavity update equations
are then the final-time marginalizations
\begin{align}
    p_{i\to (ji)}\left(\sigma_i,\sigma_j\right)=\sum_{X^t_i:\sigma_i(t)=\sigma_i}
\sum_{X^t_j:\sigma_j(t)=\sigma_j}\mu_{i\to (ji)}\left(X^t_i,X^t_j\right)
\end{align}
and the closure of the cavity update equations as master-equation-like
differential equations reads \cite{David3}
\begin{align}
\frac{d}{dt}{p}_{i\to (ij)}(\sigma_i,\sigma_j) &=  \sum_{\sigma_{\partial i \setminus j}} \Big[ r_i(\sigma_i,\sigma_{\partial i})  \prod_{ k \in \partial i\setminus j} p_{k\to (ki)}(\sigma_k \mid \sigma_i) p_{i\to (ij)}(\sigma_i, \sigma_j) \nonumber\\ 
&\qquad- r_i(-\sigma_i,\sigma_{\partial i})  \prod_{ k \in \partial i\setminus j} p_{k\to (ki)}(\sigma_k \mid -\sigma_i) p_{i\to (ij)}(-\sigma_i, \sigma_j) \Big]\nonumber \\ 
&\qquad- r_j(\sigma_j,\sigma_i) p_{i\to (ij)}(\sigma_i,\sigma_j)
+ r_j(-\sigma_j,\sigma_j) p_{i\to (ij)}(\sigma_i,-\sigma_j).
\label{eq:iCME-update}
\end{align}
In the above the conditional probabilities in the cavity are defined as
\begin{align}
p_{i\to (ij)}(\sigma_i \mid \sigma_j)=\frac{p_{i\to (ij)}(\sigma_i, \sigma_j)}{\sum_s p_{i\to (ij)}(s, \sigma_j)}.    
\end{align}
Further,
$r_i(\sigma_i,\sigma_{\partial i})$ and $r_j(\sigma_j,\sigma_i)$ are the 
defined jump rates of spins
$i$ and $j$ in the cavity graph obtained by eliminating all neighbours of $j$ except $i$.
The rate $r_i$ hence depends on all neighbours of $i$ in the original graph, including $j$,
while the rate $r_j$ only depends on $i$ and $j$. 

%\begin{eqnarray}
%\dot{p}_t(\sigma_i||X_j) &=&  \sum_{\sigma_{\partial i \setminus j}} \Big[ r_i(\sigma_i,\sigma_{\partial i}) % \prod_{ k \in \partial i\setminus j} p(\sigma_k\mid \sigma_i) p(\sigma_i||X_j) \\ \nonumber
%&-& r_i(-\sigma_i,\sigma_{\partial i})  \prod_{ k \in \partial i\setminus j} p(\sigma_k\mid -\sigma_i) %p(-\sigma_i||X_j) \Big] 
%\label{eq:weagree2}
%\end{eqnarray}

The fundamental object of the cavity output equations
are analogously the final-time marginalizations
\begin{align}
    P_i\left(\sigma_i\right)=\sum_{X^t_i:\sigma_i(t)=\sigma_i} P^t_i(X^t_i)
\end{align}
and the differential equations substituting for \eqref{eq:cavity-output-X}
are
\begin{align}
\frac{d}{dt}{P}_i(\sigma_i) 
 &=  \sum_{ \sigma_{\partial i}}\Big[ r_i(\sigma_i,\sigma_{\partial i})  \prod_{ k \in \partial i\setminus j} p_{k\to (ki)}(\sigma_k | \sigma_i) P_i(\sigma_i) \nonumber \\ 
&\quad- r_i(-\sigma_i,\sigma_{\partial i})  \prod_{ k \in \partial i\setminus j} p_{k\to (ki)}(\sigma_k | -\sigma_i) P_i(-\sigma_i) \Big]. \label{eq:iCM-output}
\end{align}

\begin{figure}[h]
\centerline{\includegraphics[width=10cm]{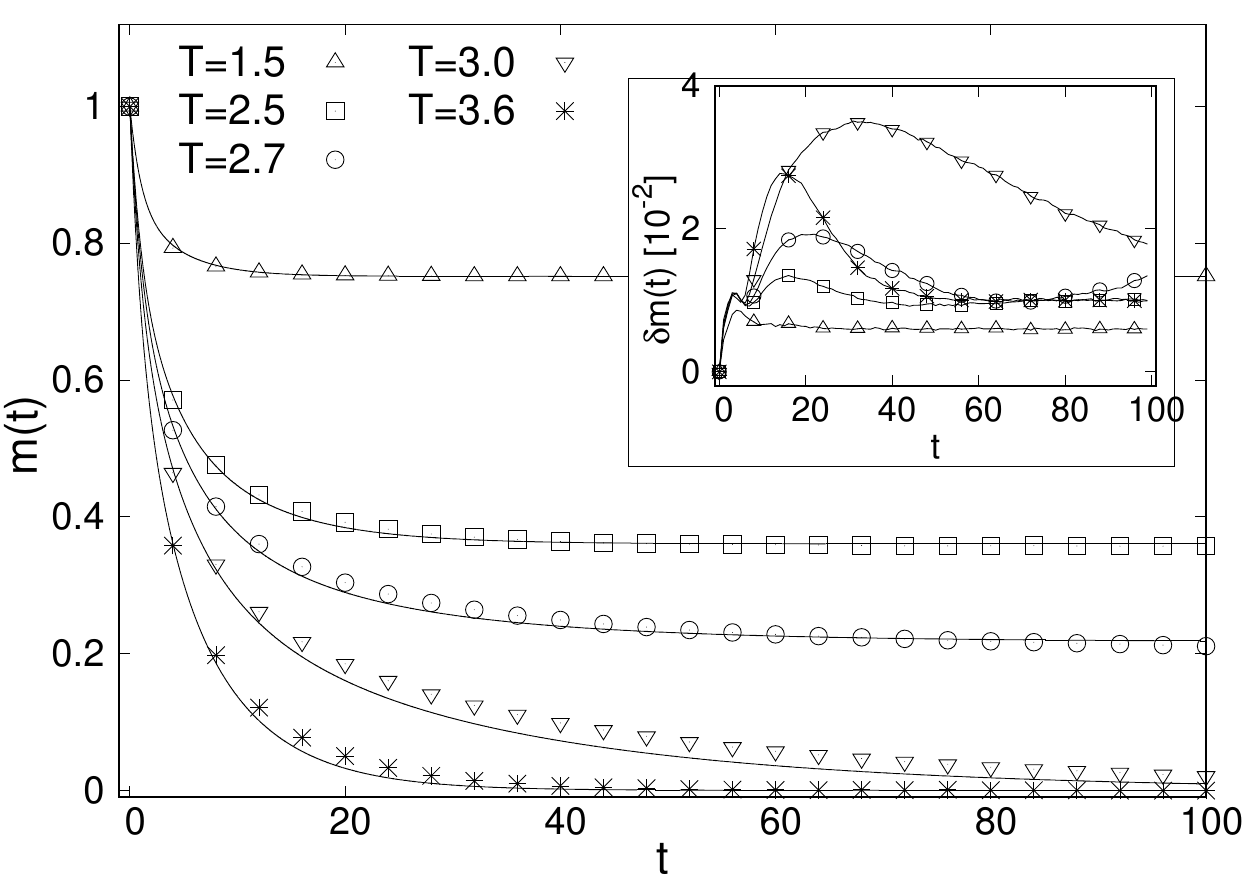}}
\caption{Continuous-time dynamics of the Ising ferromagnet in a single instance of an Erdos-Renyi graph of size $N=5000$ and average connectivity $c=3$. 
Temperature from $T=1.5$ to $T=3.6$ in standard parameterization ($J=k_B=1$) as indicated. 
The main panel shows the time evolution of the system magnetization. Points are the averages of $s=10,000$ kinetic Monte Carlo simulations of the dynamics, lines are the results from simulating \eqref{eq:iCME-update} and \eqref{eq:iCM-output}.
In all calculations, an initially fully magnetized system evolves in time in contact with a heat bath at a given temperature. The insert shows the mean square error $\delta m(t) = (N^{-1}  \sum_{i=1}^{N} (m_i^{DCAV}(t) - m_i^{MC}(t))^{2})^{1/2}$ on scale of order $10^{-2}$. Figure contributed by David Machado Perez.}
\label{fig.iCME}
\end{figure}

In Fig.~\ref{fig.iCME} we show numerical results on the kinetic Ising model
obtained using \eqref{eq:iCME-update} and \eqref{eq:iCM-output}; it can be checked 
that they improve on the earlier version of the continuous-time closure \cite{Dominguez2020}. 

%\begin{eqnarray}
%\dot{P}_t(\sigma_i) &=&  \sum_{ \sigma_{\partial i}}\Big[ r_i(\sigma_i,\sigma_{\partial i})  %P(\sigma_i,\sigma_{\partial i}) \\ \nonumber
%&-& r_i(-\sigma_i,\sigma_{\partial i}) P(-\sigma_i,\sigma_{\partial i}) \Big] \\ \nonumber
% &=&  \sum_{ \sigma_{\partial i}}\Big[ r_i(\sigma_i,\sigma_{\partial i})  \prod_{ k \in \partial i\setminus %j} p(\sigma_k\mid \sigma_i) P(\sigma_i) \\ \nonumber
%&-& r_i(-\sigma_i,\sigma_{\partial i})  \prod_{ k \in \partial i\setminus j} p(\sigma_k\mid -\sigma_i) %P(-\sigma_i) \Big] 
%\label{eq:MasterEquation}
%\end{eqnarray}
%substituting the conditional probabilities at time $t$ with cavity probabilities at time $t$.

\subsection{Problems and challenges}
\label{sec21.4.3}
We end by listing a set of open problems of a general nature. 

\vspace{5pt}
\noindent\textbf{Representation of history in dynamic cavity} \ \ %\emph{
Almost all uses of dynamic cavity, including the recent one outlined above,
have relied on some kind of Markov assumption. 
As shown in Fig.~\ref{fig.iCME} the results can be quite accurate, but
are not exact.
By analogy to other problems in
physical kinetics the Markov assumption ought to be a restrictive assumption
when spatial and temporal correlations may dominate the dynamics
of the system.  At the moment it is a clear challenge to

\vspace{5pt}
\noindent\emph{Develop efficient and general methods to handle the properties of trajectories in a compact way beyond Markov assumptions.
}

\vspace{5pt}
\noindent\textbf{Average dynamics in dynamic cavity}\ \
A great success of the cavity method as a physical theory
is that it can deal with disorder considering 
self-consistency equations of distributions over 
messages. In most cases this has been considered on the 
level of \textit{Replica Symmetry} where the self-consistency equations
take the form average or representative equations. This is the framework exploited by the
Dynamic Cavity Method, and then extended to the Cavity Master Equation for continuous time.
One approach has been developed for the average case \cite{Eduardo2}, but does not
work in many cases. The problem may be on the level of the closure, and not on a 
more fundamental level (for this, see below). Nevertheless, this remains a challenge
which we formulate as
%However, it simple cases it  is no more
%accurate then an in principle simpler dynamic mean-field method 
%developed quite some time ago \cite{Semerjian2004}, and has not hereto
%been developed  for dynamics not in detailed balance.
%We therefore formulate the resulting challenge as

\vspace{5pt}
\noindent\emph{To develop a scheme on the Replica Symmetry level to describe 
the typical properties of dynamic cavity. 
}

\vspace{5pt}
\noindent\textbf{Replica Symmetry Breaking} \ \ 
An even wider success of the cavity method has of course 
been its extension beyond Replica Symmetry, first for the 
Bethe spin glass \cite{MezardParisi2001}, and later for many
famous constraint satisfaction and  combinatorial optimization problems \cite{mezard2002analytic,mulet2002coloring,krzakala2007gibbs,decelle2011asymptotic}.
How to extend this theory to dynamics is not clear. On the technical level, iterations in 1-step Replica Symmetry Breaking
(survey propagation) are weighted by a free energy shift. As non-equilibrium
dynamics includes cyclic motion, in general it is not associated to a globally defined
free energy function.
We state this challenge as

\vspace{5pt}
\noindent \emph{Can Replica Symmetry be broken in dynamics? And can one construct a 
survey-propagation-like scheme to describe a putative 1-step Replica Symmetry Breaking
phase of dynamics?
}

\bibliographystyle{unsrt_abbvr}
\bibliography{chapter21}

\end{document}